\documentstyle[12pt]{article}
\def\lp {l_p}
\def\ls {l_s}
\def\gs {g_s}
\def\Mtilde {\widetilde{M}}
\def\lptilde {\tilde{l_p}}

\def\lstilde {\tilde{\ls}}
\def\gstilde {\tilde{\gs}}
\def\xtilde {\tilde{x}}
\def\xdoubtilde {\tilde{X}}
\def\rtilde {\widetilde{r}}
\def\Utilde {\widetilde{U}}
\def\square {\setlength{\unitlength}{1mm}\begin{picture}(3.8,3.8)
             \put(0,-0.2){\framebox(3.5,3.5){$\mbox{}$}}\end{picture}}
\def\np    { Nucl. Phys. }
\def\pr    { Phys. Rev. }
\def\pl    { Phys. Lett. }

\def\prl   { Phys. Rev. Lett. }

\def\del {\partial}

\def\be{\begin{equation}}     
\def\ee{\end{equation}}
\def\bea{\begin{eqnarray}}     
\def\eea{\end{eqnarray}}
\def\&{&\!\!\!\!\!\!\!\! &}

\newcommand{\eq}[1]{(\ref{#1})}

\def\nn{\nonumber}

\def\parbigskip        {  \par\bigskip  }
\def\parmedskip        {  \par\medskip  }

\def\parmedskipn        {  \par\medskip\noindent  }

\renewenvironment{thebibliography}{\pagebreak[3]\par\vspace{0.6em}
\begin{flushleft}{\large \bf References}\end{flushleft}
\vspace{-1.0em}

\begin{enumerate}\if@twocolumn\baselineskip=0.6em\itemsep -0.2em
\else\itemsep -0.2em\fi\labelsep 0.1em}{\end{enumerate}}

\begin{document}
\baselineskip=0.65cm





\begin{titlepage}

    \begin{normalsize}
     \begin{flushright}
                 YITP-98-85\\
                 hep-th/9812218\\
                 December 1998
     \end{flushright}
    \end{normalsize}
    \begin{LARGE}
       \vspace{1cm}
       \begin{center}
        $\mbox{AdS}_{7}/\mbox{CFT}_{6}$ Correspondence \\
          and \\
        Matrix Models of M5-Branes \\
       \end{center}
    \end{LARGE}
  \vspace{5mm}

\begin{center}
           Hidetoshi Awata 
           \footnote{E-mail address:
              awata@yukawa.kyoto-u.ac.jp}
            and Shinji Hirano
           \footnote{E-mail address:
              hirano@yukawa.kyoto-u.ac.jp}   \\
      \vspace{4mm}
        {\it Yukawa Institute for Theoretical Physics} \\
        {\it Kyoto University}\\
        {\it Sakyo-ku, Kyoto 606-8502, Japan}\\
      \vspace{1cm}

    \begin{large} ABSTRACT \end{large}
        \par
\end{center} 
\begin{quote}
 \begin{normalsize}
We study the large $N$ limit of matrix models of M5-branes, or (2,0)
six-dimensional superconformal field theories, by making use of the
Bulk/Boundary correspondence. Our emphasis is on the relation between
the near-horizon limit of branes and the light-like limit of
M-theory. In particular we discuss a conformal symmetry in the D0 + D4
system, and interpret it as a conformal symmetry in the discrete
light-cone formulation of M5-branes. We also compute two-point
functions of scalars by applying the conjecture for the AdS/CFT
correspondence to the near-horizon geometry of boosted M5-branes. We
find an expected result up to a point subtle, but irrelevant to the IR
behavior of the theory. Our analysis matches with the Seiberg and
Sen's argument of a justification for the matrix model of M-theory. 

 \end{normalsize}
\end{quote}

\end{titlepage}
\vfil\eject





\section{Introduction}

The relation between gravity and gauge theory has been explored extensively for recent a few years. There seem to be two branches in such a direction. The one is matrix models proposed by \cite{BFSS,DLCQ} for M-theory and by \cite{IKKT} for type IIB string theory as non-perturbative formulations of string theories. The other is the AdS/CFT correspondence, or more generally the Bulk/Boundary correspondence, of \cite{malda1,malda2}, and it has been elaborated in \cite{GKP,wit2pt}. 

Since both relations may originate from the $s$-$t$ channel duality, or equivalently from the UV/IR correspondence \cite{SW,PP}, of string theory,
one can expect that there is a close relation between the M(atrix) conjecture of \cite{BFSS,DLCQ} and the Maldacena conjecture of \cite{malda1,malda2}.
In fact the Seiberg-Sen limit \cite{seiberg,sen} for the matrix model of M-theory turns out to be in good harmony with the near-horizon limit of Maldacena, as discussed by Hyun and Kiem \cite{hyun}. Thus it is expected that the Bulk/Boundary correspondence is utilized as an effective tool to analyze the large $N$ limit of matrix models. Indeed in \cite{JY} Jevicki and Yoneya considered the large $N$ limit of the matrix model of M-theory from this viewpoint, and emphasized the existence of a conformal symmetry, called generalized conformal symmetry \cite{JKY}, which might play an important role in the understanding of matrix models.

In the present paper, we analyze the large $N$ limit of matrix models of M5-branes, or six-dimensional (2,0) superconformal field theories \cite{M5matrix}, by making combined use of the M(atrix) and Maldacena conjectures. Since these matrix models are conjectured to describe, not quantum gravities, but local quantum field theories, it is expected to be simpler than the matrix model of M-theory for testing the M(atrix) conjecture ({\it e.g.} covariance {\it etc.}). Thus they may be helpful for the better understanding of the matrix model of M-theory.

We give a brief review of matrix models of M5-branes, or six-dimensional (2,0) superconformal field theories in sect. \ref{sec:prelim}. Then we can find that the geometry which we should consider is the near-horizon geometry of D0 + D4 bound states. We analyze it in sect. \ref{sec:branesol}, and argue in particular a generalized conformal symmetry of the type discussed in \cite{JY}. We also give the near-horizon geometry of \lq\lq boosted" M5-branes, which is the M-theory counterpart of the near-horizon geometry of D0 + D4 bound states. A certain effective action of a particle is calculated in two ways; one way from \lq matrix models' and the other way from \lq the discrete light-cone quantization (or DLCQ) of M-theory'. We confirm agreement of both ways.

This agreement is naturally explained from the argument in sect. \ref{sec:limits}. We study in detail a relation between the near-horizon limit and the Seiberg-Sen limit in the case of the DLCQ of M5-branes. It should be stressed that, assuming the Maldacena's conjecture for our problem, we can justify the Seiberg and Sen's argument more strongly than in the case of the DLCQ of M-theory. This is because we take the decoupling limit for \lq\lq boosted" M5-branes (corresponding to M-theory) as well as for D0 + D4 bound states (corresponding to matrix model), and thus it is sufficient in the large $N$ limit only to consider the classical supergravities on their near-horizon geometries. Furthermore, since in our case \lq M-theory' itself has a conformal symmetry, we can clarify an eleven-dimensional origin of the generalized conformal symmetry by considering an alternative representation for the one given in sect. \ref{sec:branesol}. 

In sect. \ref{sec:2ptfunc} we calculate two-point functions of the DLCQ of (2,0) superconformal theories by applying the conjecture of \cite{GKP,wit2pt} for the AdS/CFT correspondence to our problem, and find an expected result up to a subtle correction which can be discarded concerning the IR behavior of the theory. 

Sect. \ref{sec:discuss} is devoted to conclusions and discussions.
In Appendix A we extend some results in sect. \ref{sec:branesol} to more general cases.
Finally in Appendix B we give some technical details to solve the differential equation for scalars in sect. \ref{sec:2ptfunc}.




\section{Preliminaries}\label{sec:prelim}

We begin with a brief review of matrix models of M5-branes, or (2,0) superconformal field theories, \cite{M5matrix,ABS}.

Let us consider M-theory in the background of longitudinal M5-branes. According to the M(atrix) conjecture of \cite{BFSS,DLCQ}, its DLCQ description is expected to be a quantum mechanics of the D0 + D4 bound states \cite{berkdoug}.
It is a $U(N)$ super Yang-Mills quantum mechanics with an adjoint hypermultiplet and $k$ fundamental hypermultiplets. D0-branes moving away from D4-branes describe supergravitons in this background. 

Now for M5-branes to decouple from the bulk, we take the limit that the eleven dimensional Planck length $\lp$ goes to zero. Thus in this limit D0-branes are confined in the D4-branes.
In terms of the super Yang-Mills quantum mechanics, this limit corresponds to $g_{YM}\to\infty$, and the Higgs branch decouples from the Coulomb branch. Furthermore it is known that the Higgs branch is equivalent to the moduli space of $N$ $U(k)$ Yang-Mills instantons. This is intuitively because a D0-brane can be considered as the zero size limit of a 4D Yang-Mills instanton on the 5D worldvolume of D4-branes. In fact the Higgs branch gives the ADHM construction of instantons,
\begin{eqnarray}
&&[X, X^{\dagger}] - [\tilde{X}, \tilde{X}^{\dagger}] 
 + q_i q_i^{\dagger} -\tilde{q}^{i\dagger}\tilde{q}^i =0, \nn\\
&&[X, \tilde{X}] + q_i \tilde{q}^i =0. \nn
\label{eqn:ADHM}
\end{eqnarray}
Thus we are led to the conjecture that the DLCQ of M5-branes, or (2,0) superconformal theories, can be described by a quantum mechanics on certain instanton moduli spaces \cite{M5matrix}.

\input epsf
\begin{figure}[h]
 \begin{center}
  \leavevmode\hbox{\epsffile{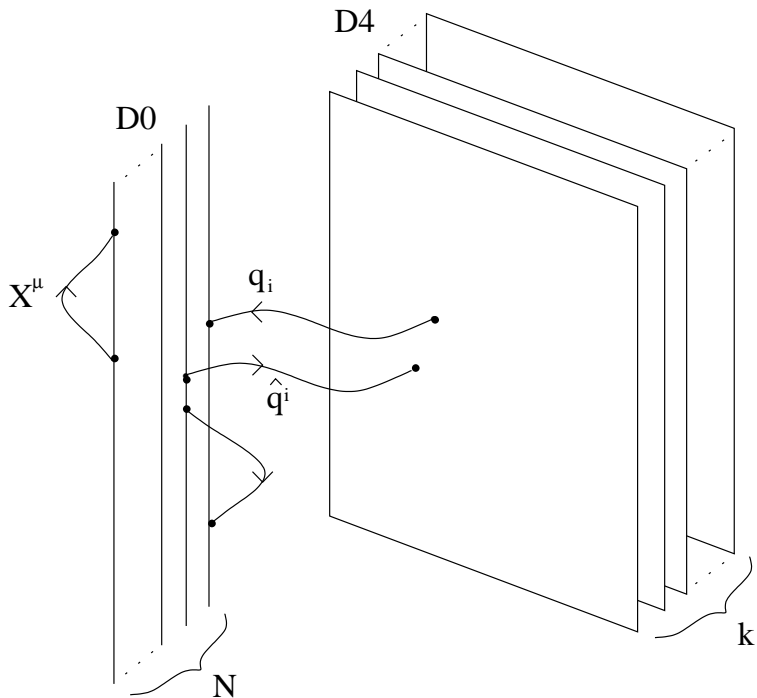}}\\[3mm]
 \caption{A D0 + D4 bound state: An adjoint hypermultiplet (two complex scalars $X=X^0 +iX^1, \tilde{X}=X^2 +iX^3$) comes from 0-0 strings, while the fundamental hypermultiplets ($q_i, \tilde{q}^i, \,\, i=1,\cdots, k$) are supplied from 0-4 and 4-0 strings.}
 \end{center} 
\label{fig:D0D4}
\end{figure}



\section{Brane solutions in the near-horizon limit}\label{sec:branesol}

Next we consider the supergravity solution which
corresponds to the large $N$ limit of matrix models of M5-branes, or (2,0)
superconformal theories. It is expected to be the near-horizon geometry of the D0 + D4 bound states, by taking into account the argument in the previous section and the Maldacena's conjecture of \cite{malda1,malda2}. This geometry enjoys a generalized conformal symmetry proposed in \cite{JY,JKY}, and just in the same way as the work of Jevicki and Yoneya \cite{JY}, it determines
the probe D0-brane action in the background of a source with a large
number of D0- and D4-branes. We also consider the M-theory counterpart of the D0 + D4 system, and it turns out to be the near-horizon geometry of \lq\lq boosted" M5-branes constructed in \cite{CT}. We calculate a particle action with fixed light-cone momentum in this background, and find agreement with the probe D0-brane action determined by the generalized conformal symmetry.

\subsection{D0 + D4 bound states} 

Let us consider the type IIA theory 
in the background of $Q_0$ D0-branes and $Q_4$ D4-branes. 
Its low energy behavior is described by the supergravity on the following geometry:
\bea
ds_{10}^2 &=&
-{dt^2 \over \sqrt{H_0 H_4} }
+\sqrt{{H_0 \over H_4}} dx_{\parallel}^2 
+\sqrt{H_0 H_4} (dr^2 + r^2 d\Omega_4^2),
\cr
e^{-2\phi} &=& 
g_s^{-2} H_0^{-{3\over 2}} H_4^{{1\over 2}},
\qquad\qquad
H_i = 1+{g_s l_s^3 Q_i \over r^3},
\cr
A_0 &=& 
g_s^{-1} (1-H_0^{-1}),
\label{e:D0D4}
\eea
where $\phi$ is a dilaton and $A_0$ is a RR vector potential.
$x_{\parallel}$ denotes the spatial coordinates $x^i$ ($i=1,2,3,4$) parallel to D4-branes, and $r$ and $\Omega_4$ denote the polar coordinates transverse to D4-branes. D0-branes sit at a point in $x_{\parallel}$-directions on D4-branes. The parameters $l_s$ and $g_s \equiv e^{\phi(r=\infty)}$ are the string length and the string coupling constant, respectively.

We now take the near-horizon limit $r\rightarrow 0$ and 
$l_s\rightarrow 0$ (and also $g_s\to 0$ and $x_{\parallel}\to 0$) as in \cite{malda1,malda2}, keeping fixed the following three quantities:
\be
U\equiv {r\over l_s^2}, \qquad
{x_{\parallel}\over l_s},\qquad
{g_s \over l_s}.
\ee
Here $U$ corresponds to the energy of open strings stretched between D-branes.
$\gs/\ls$ coincides with the Yang-Mills coupling constant in three dimensions, but we are not sure its meaning in our context.
In this limit, the above solution becomes
\bea
ds_{10}^2 &=&
l_s^2\left[
- {U^2\over H} dt^2 
+ \sqrt{{Q_0 \over Q_4}} \left({dx_{\parallel}\over l_s}\right)^2 
+ H \left({dU^2\over U^2} + d\Omega_4^2\right)
\right],
\cr
e^{-2\phi} &=& 
Q_4^2 H^{-3},
\qquad\qquad
H = {g_s\over l_s} {\sqrt{Q_0 Q_4}\over U},
\cr
A_0 &=& 
- l_s Q_4 U H^{-2},
\label{e:NearD0D4}
\eea
where we have gauged away a constant part of $A_0$.
If $H$ in the above equation were constant, the spacetime would be
$AdS_2 \times S^4 \times {\bf R}^4$ with radius $l_s \sqrt H$.\footnote{Although this metric is not $AdS_2$, it can be written as $\{(\mbox{Weyl factor})(AdS_2 \times S^4)\}\times {\bf R}^4$.}

In order to trust the supergravity solution (\ref{e:NearD0D4}), the curvature 
$R \propto U/(g_s l_s \sqrt{Q_0 Q_4}) = l_s^{-2} H^{-1}$
and the effective string coupling $e^{\phi}$ must be small.
These conditions are given by $R\ll l_s^{-2}$ and $e^\phi \ll 1$, and thus we have
\be
{g_s\over l_s}Q_0^{1\over 2} Q_4^{-{1\over 6}}
\ll U \ll
{g_s\over l_s}(Q_0 Q_4)^{1\over 2}.
\ee
Hence, for large $Q_0$ and $Q_4$, we have wide range to trust the supergravity solution.

Similarly to \cite{JY},
it is easy to show that this metric and the dilaton are invariant under 
the transformations,
the time translation $\delta_H$, 
the dilatation $\delta_D$ and 
the special conformal transformation $\delta_K$:
\be
\begin{array}{rcl}
\delta_H\, t   &=& 1, \\
\delta_H\, U   &=& 0, \\
\delta_H\, g_s &=& 0, 
\end{array}
\qquad
\begin{array}{rcl}
\delta_D\, t   &=& -t, \\
\delta_D\, U   &=& U, \\
\delta_D\, g_s &=& g_s, 
\end{array}
\qquad
\begin{array}{rcl}
\delta_K\, t   &=& -t^2- \left(g_s\over l_s\right)^2{Q_0 Q_4\over 2 U^4}, \\
\delta_K\, U   &=& 2tU, \\
\delta_K\, g_s &=& 2tg_s. 
\end{array}
\label{e:D0D4translation}
\ee
They form an $SU(1,1)$ algebra,
\be
[\delta_D,\delta_H] =  \delta_H,\qquad
[\delta_D,\delta_K] = -\delta_K,\qquad
[\delta_H,\delta_K] = 2\delta_D.
\ee
Note that $H$ in eq. \eq{e:NearD0D4} 
is invariant under these transformations.
In the solution \eq{e:NearD0D4} 
and the $SU(1,1)$ transformation \eq{e:D0D4translation}, 
$g_s$ and $Q_0$ appear only through the combination $g_s \sqrt Q_0$ 
except for the $dx_{\parallel}^2$ part of the metric.
Therefore, if we restrict ourselves at the origin $x_{\parallel}=0$,
the transformation laws for $g_s$ can be replaced with 
those for $\sqrt Q_0$.
In the next section, we will utilize this property and give an alternative representation of this transformation. In particular we will discuss its meaning from the eleven-dimensional point of view.

Next let us consider the scattering of a probe D0-brane 
in the background of the source D0 + D4 system.
If we consider only the motion along the radial direction $U$,
the effective action can be determined by the above $SU(1,1)$ symmetry.
Actually the invariance under the translation $\delta_H$, the dilatation $\delta_D$ 
and the time inversion
restricts the effective action into the form
\be
S_{eff} = \int dt \,
F\left(H,{\dot U^2\over U^4}\right) \, U, 
\ee
with an arbitrary function $F$.
Here we assumed that the effective action is independent of 
the time derivative of the string coupling constant $g_s$.
Then, from the invariance under the special conformal transformation,
we can fix the form of the effective action into 
\bea
S_{eff} &=& 
\int dt\, 
\left(\sum_{n\geq 1}b_n w^n\right)\,
f(H) \, U
,
\cr
b_{n+1} &=& {2n-1 \over 2n+4} b_n,
\qquad
w = H^2 {\dot U^2 \over U^4}
= Q_0 Q_4 \left({g_s\over l_s}\right)^2{\dot U^2\over U^6},
\eea 
with an arbitrary function $f(H)$.
Thus the effective action of a probe D0-brane moving away from the source D0 + D4 -branes is determined as 
\be
S_{eff} =
\int dt\, 
\left(1-\sqrt{1-w}\right)\,
f(H) \, U.
\label{e:probe}
\ee
This effective action would be that of matrix models of M5-branes in the large $Q_0$ (and $Q_4$) limit.
Next we will show that it coincides with the one derived from the DLCQ of M-theory.


\subsection{M5 branes boosted along the longitudinal direction}


In the case of pure D0-branes \cite{JY}, 
they found agreement between the effective action for a probe D0-brane
determined by the generalized conformal symmetry (corresponding to the matrix model) and a particle action \cite{BBPT} with fixed light-cone momentum in the plane-fronted wave background, i.e., the Aichelburg-Sexl metric (corresponding to the DLCQ of M-theory).
From the viewpoint of the Bulk/Boundary correspondence, the DLCQ of (large $N$) M5-branes, or (2,0) superconformal theories, is expected to be described by the near-horizon limit of ``longitudinally boosted'' M5-branes.
Although the M5-brane solution is isometric in the longitudinal directions and thus it is boost invariant, one can define ``longitudinally boosted'' M5-branes as an extreme limit of longitudinally boosted {\it non-extreme} M5-branes.


Let us start from a non-extreme, or non-BPS, $N$ coincident M5-brane solution \cite{CT} with vanishing three-form
\bea
ds_{M5}^2 &=&
H^{-{1\over 3}}
\left( -f\, dt^2 + (dx^{11})^2 + dx_{\parallel}^2 \right)
+ H^{{2\over 3}} (f^{-1}dr^2 + r^2 d\Omega_4^2),
\cr
H &=& 1+{ \pi N l_p^3 \over r^3},\qquad
f  =  1-{\mu l_p^3 \over r^3},
\label{e:nonextremeM5}
\eea
which is a Schwarzschild type deformation of the extreme solution
with a non-extremality parameter $\mu$, such that
$\mu \rightarrow 0$ corresponds to the BPS-saturated limit.
To compare with the matrix model, or the D0 + D4 system,
let us compactify along a longitudinal direction, say $x^{11}$,
with radius $\widetilde R$ and 
boost it along that direction by
\be
\left[
\matrix{t'\cr {x^{11}}'}
\right]
=
\left[
\matrix{\cosh\beta & \sinh\beta \cr
        \sinh\beta & \cosh\beta}
\right]
\left[
\matrix{t\cr x^{11}}
\right].
\ee
Taking the extreme limit $\mu\rightarrow 0$ and 
boosting infinitely $\beta\rightarrow\infty$ 
with fixed $\widetilde Q\equiv\mu e^\beta$,
the first two terms in \eq{e:nonextremeM5}
map to
\bea
-f\,dt^2 + (dx^{11})^2 
&=& 
-(dt')^2  + (d{x^{11}}')^2 + 
{\mu l_p^3 \over r^3}(\cosh\beta\,dt - \sinh\beta\,dx^{11})^2,\cr
&\rightarrow&
dx^+ dx^- + {\widetilde Ql_p^3 \over r^3} (dx^-)^2,
\eea
with the light-cone coordinates
$x^{\pm}\equiv {x^{11}}' \pm t'$.
Then we obtain
the longitudinally boosted $N$ coincident M5-brane metric 
\be
ds_{M5}^2 =
H^{-{1\over 3}}
\left( dx^+ dx^- + {\widetilde Q l_p^3\over r^3} (dx^-)^2 + 
dx_{\parallel}^2 \right)
+ H^{{2\over 3}} (dr^2 + r^2 d\Omega_4^2).
\label{e:M5}
\ee
Under this infinite boost,
the space-like compactification along $x^{11}$ direction
is changed to the almost light-like compactification 
with radius $\widetilde R e^\beta$.
Therefore, in order to have a finite light-like radius,
the space-like circle must be shrunk to a point, 
i.e., $\widetilde R\rightarrow 0$, as discussed in \cite{seiberg,sen}.

In the near-horizon limit, $r\to 0$ and $l_p\rightarrow 0$ with fixed
$U_{M5}^2 \equiv {r/l_p^3}$, the solution \eq{e:M5} becomes
\be
ds_{M5}^2 =
{l_p^2 \over (\pi N)^{1\over 3}} \left[
U_{M5}^2 dx^+ dx^-
+ {Q\over R^6}{(dx^-)^2\over U_{M5}^4} 
+ U_{M5}^2 dx_{\parallel}^2
+\pi N \left(4 {dU_{M5}^2\over U_{M5}^2} + d\Omega_4^2 \right)
\right].
\label{e:NearM5}
\ee
Here we introduced a new boost parameter $Q$ and 
the finite light-like radius $R$ by
$Q/R^6 \equiv \widetilde Q/l_p^6$.
Note that the light-like radius $R$ is the only finite dimensionful parameter of this system.


If the background metric 
\be
ds_{11}^2 = 
g_{++} (dx^+)^2 + 2 g_{+-} dx^+ dx^- + g_{--} (dx^-)^2
+ g_{ij}dx^i dx^j, 
\ee
with $x^{\pm} = x^{11} \pm x^0$ and $i,j=1,2,\cdots,9$, 
is independent of $x^-$,
then the light-like momentum $p_- = \partial {\cal L}/\partial \dot x^-$ 
is a constant in time $\tau\equiv x^+/2$,
and the dynamics is well described by the Routhian \cite{BBPT},
\bea
-{\cal R} &=& 
-m \sqrt{-g_{\mu\nu} \dot x^\mu \dot x^\nu}
- p_- \dot x^- \vert_{m\rightarrow 0},
\cr
&=& p_- {2 g_{+-}\over g_{--}} 
\left(1- \sqrt{
1 - {g_{--}\over (2 g_{+-})^2} (4g_{++} + g_{ij} \dot x^i \dot x^j) 
}\right).
\eea
This is a Hamiltonian for $x_-$ and a Lagrangian for the other variables.
In our case, 
$p_- = 1/R$ and
if we restrict ourselves to the motion along the radial direction $U$, i.e., 
$\dot x_{\parallel} = \dot \Omega_4 = 0$, then we have
\be
d\tau\, {\cal R} =
d\tau\, {R^5 U_{M5}^6\over Q}
\left(1-\sqrt{1-{4\pi N Q\over R^6 U_{M5}^{10} }\dot U_{M5}^2 }\right).
\label{e:M5Routhian}
\ee

The relations between the variables in the D0 + D4 and the M5 systems 
will be naturally explained in the next section, and the result is given by
\be
\left(U,\, 2t,\, {x_{\parallel}\over l_s}\,;\,
Q_0,\, Q_4,\, {l_s\over g_s}\right)_{D0D4} 
=
\left(R U_{M5}^2,\, x^+,\, {x_{\parallel}\over R}\,;\,
Q,\, \pi N,\, R \right)_{M5}.
\label{e:D0D4andM5}
\ee
Under this identification, 
the Routhian \eq{e:M5Routhian} coincides with the probe action \eq{e:probe} 
with $f(H) = H^{-2} Q^{-1}$.
At first sight, the correspondence of 
the time $t$ in ten dimensions and 
the light-cone time $x^+$ in eleven dimensions
 looks somewhat mysterious, but as we will discuss in the next section,
it turns out to orginate from the noncommutativity of two procedures, i.e., to take the near-horizon limit and to uplift the ten-dimensional metric.




\section{The light-like limit and the near-horizon limit}\label{sec:limits}

As discussed in the work of Hyun and Kiem \cite{hyun}, the scaling
limits given by Seiberg \cite{seiberg} and Sen \cite{sen} are in good
harmony with the near-horizon limits in the AdS/CFT
correspondence. Here we will argue the relationship between these two
kinds of limits in detail, in the case of the light-cone description
of M5-branes, or (2,0) superconformal field theories. 
\parmedskip
We start with a brief review of the arguments given in \cite{seiberg} and \cite{sen}. For later use, it is enough to discuss only the case of compactification on a circle.

Now let us compactify M-theory on a light-like circle,
\begin{equation}
x^{-} \simeq x^{-} + 2\pi R,
\label{eqn:LL}
\end{equation}
where we defined the light-like coordinates $x^{\pm} = {1 \over \sqrt{2}}(x^{11} \pm t)$.

The light-like compactification (\ref{eqn:LL}) may be viewed as an infinite boost limit (along the $x^{11}$-direction) of the compactification on a space-like circle,
\begin{equation}
x^{11} \simeq x^{11} + 2\pi \widetilde{R}.
\label{eqn:SL}
\end{equation}
In fact one can easily find that it is obtained as an $\widetilde{R} \to 0$ limit of a large boost with parameter $\beta = R/\sqrt{R^2 + 2\widetilde{R}^2}$ \cite{seiberg}. We should, however, note that it is not trivial whether this limit really exists or not in general \cite{HelPol}.\footnote{The author in \cite{Bilal} argued some evidences for the existence of the light-like limit of M-theory.}

Assuming the Lorenz invariance and the existence of the above light-like limit for M-theory, we can map M-theory on a space-like circle (\ref{eqn:SL}) to another M-theory on a light-like circle (\ref{eqn:LL}) by an infinite boost along the longitudinal direction and by rescaling the parameters of the theory. Let us call the former theory $\Mtilde$ with Planck length $\lptilde$, and the latter $M$ with $\lp$. Now since we are interested in a sector with fixed longitudinal momentum $P_{11} = N/\widetilde{R}$ for $\Mtilde$, and fixed light-cone momentum $P_{-} = N/R$ for $M$, the energy scale of the theory is $\widetilde{R}/{\lptilde}^2$ for $\Mtilde$ (for small $\widetilde{R}$ or in the IMF) and $R/{\lp}^2$ for $M$, respectively. Therefore the Hamiltonian of these two theories match up in the scaling limit,
\begin{eqnarray}
&\widetilde{R} \to 0,& \nn\\
\label{eqn:doublelim} 
&\lp \to 0,& \\
&\widetilde{R}/{\lptilde}^2 = R/{\lp}^2 = \mbox{fixed}.& \nn
\end{eqnarray}
The transverse directions are not affected by a longitudinal boost, and thus we should relate the transverse coordinates $\xtilde^i$ of $\Mtilde$ to those $x^i$ of $M$ via
\begin{equation}
\xtilde^i /\lptilde = x^i /\lp.
\label{eqn:transverse}
\end{equation}
From the scaling limit (\ref{eqn:doublelim}), $\Mtilde$ reduces to type IIA theory, denoted by $\widetilde{IIA}$, with the string coupling $\gstilde$ and the string length $\lstilde$,
\begin{eqnarray}
&\gstilde = (\widetilde{R}/\lptilde)^{3/2} = \widetilde{R}^{3/4}(R/\lp^2)^{3/4} \to 0,& \nn\\
&\lstilde^2 = (\widetilde{R}/\lptilde^3)^{-1} = \widetilde{R}^{1/2}(R/\lp^2)^{-3/2} \to 0.&
\label{eqn:IIAparameters}
\end{eqnarray} 
Thus the DLCQ of M-theory is described by the low energy effective theory of D0-branes in type IIA theory.
\parbigskip
Now it is rather easy to see the compatibility of the above scaling limit with the near-horizon limit in \cite{malda1,malda2}. From the relation (\ref{eqn:transverse}) of transverse coordinates, we have for radial coordinates,
\begin{equation}
\rtilde/\lptilde = r/\lp.
\label{eqn:radial}
\end{equation}
This leads us to the near-horizon limit of Maldacena,
\begin{eqnarray}
&\lstilde \to 0,& \nn\\
\label{eqn:maldalim}
&\rtilde \to 0,& \\
&\Utilde \equiv \rtilde/\lstilde^2 = r(R/\lp^3) = \mbox{fixed}.& \nn\\
&g_{YM}^2 = \gstilde/\lstilde^3 = (\widetilde{R}/\lptilde^2)^3 = (R/\lp^2)^3 
= \mbox{fixed}& \nn
\end{eqnarray}
Thus the scaling limit of Seiberg and Sen is compatible with the near-horizon limit employed for D0-branes in the SUGRA/SYM correspondence \cite{malda2}.
%



\subsection{D0 + D4 bound states and \lq\lq boosted" M5-branes}

\begin{figure}[hbt]
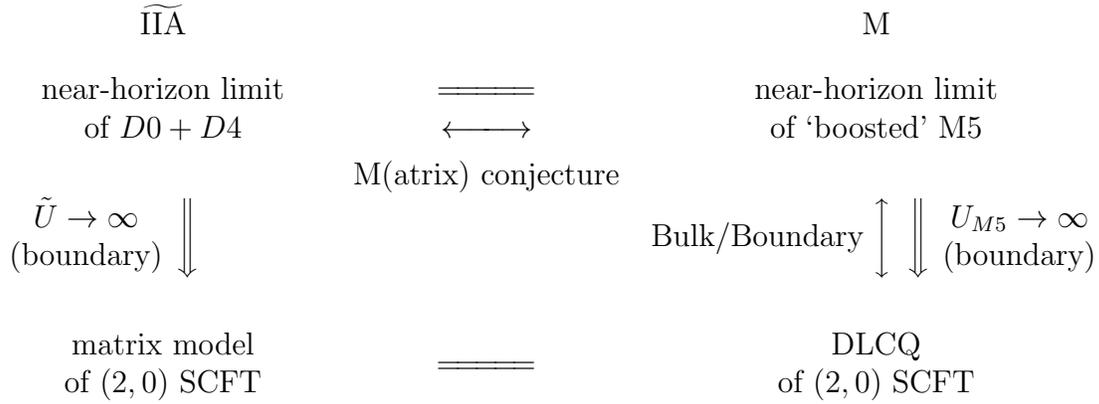

$$
\matrix{
  \vphantom{A}
&
&
\cr
  \widetilde{\rm IIA}
&
& {\rm M}
\cr
&
&
\cr
  \matrix{
          \mbox{near-horizon limit} \cr
          \mbox{of $D0+D4$} 
          } 
& \matrix{
          =\!=\!=\!=\!=  \cr
          \longleftarrow\!\!\longrightarrow 
          }
& \matrix{
          \mbox{near-horizon limit} \cr
          \mbox{of `boosted' M5}
          } 
\cr
  \vphantom{\tilde{A^A\over A} }
& \mbox{M(atrix) conjecture}
&
\cr
  \matrix{
          \tilde U \rightarrow \infty \cr
          \mbox{(boundary)}
          }\;
  \Bigg\Downarrow 
  \hphantom{ABCDE}
&  
& \mbox{Bulk/Boundary}\;
  \Bigg\updownarrow \;\; \Bigg\Downarrow \;
  \matrix{
          U_{M5} \rightarrow \infty \cr
          \mbox{(boundary)}
          }
\cr
  \vphantom{\tilde{A^A\over A}}
&
&
\cr
  \matrix{
          \mbox{matrix model} \cr
          \mbox{of $(2,0)$ SCFT}
          }
&  =\!=\!=\!=\!=
& \matrix{
          \mbox{DLCQ} \cr
          \mbox{of $(2,0)$ SCFT}
          }
\cr
  \vphantom{A}
&
&
\cr
}
$$
\caption{The horizontal direction corresponds to the M(atrix) conjecture, while the vertical direction corresponds to the Maldacena conjecture. We discuss the matrix model and the DLCQ of M5-branes in the large N limit by making use of the Bulk/Boundary correspondence.}
\label{fig:diagram}
\end{figure}

Next we turn to the analysis of the large $N$ limit of matrix models of M5-branes, and develop further the relation between two kinds of limits discussed above.
It is summarized by a diagram in fig. \ref{fig:diagram}.

Let us recall the near-horizon geometry of the D0 + D4 system,
\begin{eqnarray}
ds_{10}^2 &=& \lstilde^2\left[-\left({\lstilde \over \gstilde}\right){\Utilde^3 \over \sqrt{Q_0Q_4}}dt^2 + \left({\gstilde \over \lstilde}\right){\sqrt{Q_0Q_4} \over \Utilde^3}(d\Utilde^2 + \Utilde^2 d\Omega_4^2)\right. \nn\\
          && + \left. \sqrt{{Q_0 \over Q_4}}\left\{(d\xdoubtilde^1)^2 
            + (d\xdoubtilde^2)^2 + (d\xdoubtilde^3)^2 + (d\xdoubtilde^4)^2 
            \right\}\right], \nn\\
\label{eqn:D0D4}
e^{-2\phi} &=& \left({\lstilde \over \gstilde}\right)^3 Q_0^{-{3 \over 2}}
               Q_4^{1 \over 2}\Utilde^3,
\qquad A_0 = -\left({\lstilde \over \gstilde}\right)^2 Q_0^{-1}\Utilde^3 
              \lstilde,
\end{eqnarray}
where $\Utilde = \rtilde/\lstilde^2$, $\xdoubtilde^i = \xtilde^i/\lstilde$ ($i=1,2,3,4$), and we are taking the limit 
\begin{eqnarray}
&&\lstilde \to 0, \qquad \gstilde \to 0, \nn\\
\label{eqn:D0D4limit}
&&\rtilde \to 0, \qquad \xtilde^i \to 0, \\
&&\Utilde = \mbox{fixed}, \quad {\gstilde \over \lstilde} = \mbox{fixed},
 \quad \xdoubtilde^i = \mbox{fixed}. \nn 
\end{eqnarray}
From the viewpoint of the Seiberg and Sen's argument, this system
resides on $\widetilde{IIA}$-theory. As we will show below, the $M$-theory
counterpart of this system is the near-horizon geometry of M5-branes
boosted along the longitudinal direction, as already discussed in the previous section.

Now we reconsider the near-horizon limit (\ref{eqn:D0D4limit}) in view of the Seiberg and Sen's argument. First from the relations (\ref{eqn:IIAparameters}) for the string coupling and the string length, we find
\begin{equation}
{\gstilde \over \lstilde} = \widetilde{R}^{1/2}(R/\lp^2)^{3/2}.
\label{eqn:gsoverls}
\end{equation}
In order to keep $\gstilde/\lstilde$ fixed, we take the decoupling limit 
\begin{equation}
\lp = \widetilde{R}^{1/6}R^{5/6} \to 0.
\label{eqn:decouple}
\end{equation}
Here the dependence on $R$ is determined purely on dimensional grounds. We have fixed the constant of proportionality to be $1$ for brevity. As a result, we have
\begin{equation}
{\gstilde \over \lstilde} = {1 \over R},
\label{eqn:gsovls}
\end{equation}
which is easily understood from the fact that there is now only one dimensionful constant $R$ in this scaling limit.

We should emphasize that the Yang-Mills coupling of the quantum mechanics for the D0 + D4 system \cite{berkdoug} goes to infinity,
\begin{equation}
g_{YM}^2 \propto \widetilde{R}^{-1} \to \infty,
\end{equation}
as opposed to the pure D0-brane case with finite Planck length $\lp$.
Thus, in this limit, the Coulomb branch decouples from the Higgs branch . In $M$-theory language, M5-branes decouple from the bulk, leaving a six-dimensional field theory, without gravity, on the five-brane worldvolume. This matches with the observation for matrix models of (2,0) superconformal field theories \cite{M5matrix}. 

Next let us look at the radial coordinate $\Utilde$ and transverse coordinates $\xdoubtilde^i$. As is given in eq. (\ref{eqn:maldalim}), we have for $\Utilde$,
\begin{equation}
\Utilde = R(r/\lp^3),
\label{eqn:utilde}
\end{equation}
and for $\xdoubtilde^i$ from eq. (\ref{eqn:transverse}),
\begin{equation}
\xdoubtilde^i = {\xtilde^i \over \lstilde} = x^iR^{1/4}(\widetilde{R}/\lp^6)^{1/4}.
\label{eqn:xdoubtilde}
\end{equation}
Note that the combination $r/\lp^3$ appeared in eq. (\ref{eqn:utilde}) has mass dimension 2 and is nothing other than the variable $U_{M5}^2$ for \lq\lq boosted" M5-branes given in the previous section. In general cases, all the parameters of $M$-theory are supposed to be finite, and so is the radial variable $\Utilde$ of $\widetilde{IIA}$-theory, as already indicated in eq. (\ref{eqn:maldalim}). In this case, however, as we are considering the decoupling limit of $M$-theory, we have to take an $r \to 0$ limit in order to keep $\Utilde$ fixed. Hence we are led to take the near-horizon limit in $M$-theory as well as in type $\widetilde{IIA}$-theory:
\begin{eqnarray}
&l_p = \widetilde{R}^{1/6}R^{5/6} \to 0,& \nn\\
\label{eqn:M5NH}
&r \to 0,& \\
&U_{M5}^2 = r/l_p^3 = \mbox{fixed}\;\; (= R^{-1}\Utilde).& \nn
\end{eqnarray}
Note also that the transverse coordinates $\xdoubtilde^i$ in type $\widetilde{IIA}$-theory are finite, since eq. (\ref{eqn:xdoubtilde}), in the decoupling limit (\ref{eqn:decouple}), deduces to
\begin{equation}
\xdoubtilde^i = R^{-1}x^i.
\label{eqn:xfinite}
\end{equation}
\parmedskip
Applying the Seiberg and Sen's argument to (2,0) superconfomal field
theories, it must be shown that the DLCQ Hamiltonian of M5-branes in
$M$-theory is equal to the Hamiltonian of D0-branes moving on
D4-branes in type $\widetilde{IIA}$-theory. In our context this
statement may be equivalent to show that the supergravity action in
the near-horizon geometry of M5-branes, compactified on a light-like
circle, with fixed light-cone momentum coincides with the one in the
near-horizon geometry of the D0 + D4 bound states.  
 
To show this, it is useful to uplift the near-horizon geometry of the D0 + D4 solution to eleven dimensions. In general the uplifted metric is given by \cite{variousdim}
\begin{equation}
ds_{11}^2 = e^{-{2 \over 3}\phi}ds_{10}^2 + e^{{4 \over 3}\phi}
           (dx^{11} - \sum_{m = 0}^{9}A_m dx^m)^2.
\label{eqn:upgeneral}
\end{equation}
In the present case, this amounts to 
\begin{eqnarray}
ds_{11}^2 &=& \lstilde^2\left[ 2{\Utilde \over RQ_4^{1/3}}d\tau dx^{-} 
            + {Q_0 Q_4^{-1/3} \over R^2 \Utilde^2}(dx^{-})^2
            + {Q_4^{2/3} \over \Utilde^2}\left(d\Utilde^2 
            + \Utilde^2 d\Omega_4^2
              \right)\right. \nn\\
\label{eqn:upD0D4}
   && \quad\qquad\left. + {R\Utilde \over Q_4^{1/3}}
      \left\{(d\xdoubtilde^1)^2 + (d\xdoubtilde^2)^2
           + (d\xdoubtilde^3)^2 + (d\xdoubtilde^4)^2 \right\}
      \right],
\end{eqnarray}
where we have defined $\tau = Rt$, $x^{-} = x^{11}/\lstilde$, and used the relation, $\lstilde/\gstilde = R$, in the scaling limit discussed above.

\noindent
Using the relations (\ref{eqn:M5NH}) and (\ref{eqn:xfinite}) between the coordinates in $\widetilde{IIA}$- and $M$-theories, we can rewrite the eleven dimensional metric to
\begin{eqnarray}
ds_{11}^2 &=& \lstilde^2\left[ 2{U_{M5}^2 \over Q_4^{1/3}}d\tau dx^{-} 
            + {Q_0 Q_4^{-1/3} \over R^4U_{M5}^4}(dx^{-})^2
            + {Q_4^{2/3} \over U_{M5}^2}\left(4dU_{M5}^2 + U_{M5}^2 d\Omega_4^2
              \right)\right. \nn\\ 
\label{eqn:upM5}
   && \quad\qquad\left. + {U_{M5}^2 \over Q_4^{1/3}}
      \left\{(dx^1)^2 + (dx^2)^2 + (dx^3)^2 + (dx^4)^2 \right\}
      \right].
\end{eqnarray}
This is exactly the metric (\ref{e:NearM5}) of \lq\lq longitudinally
boosted'' M5-branes given in the previous section.
Note that the uplifted metric is written in terms of only the
quantities in $M$-theory except for the overall coefficient
$\lstilde^2$, although it is originally described by only the
quantities in $\widetilde{IIA}$-theory. Since the overall coefficient
$\lstilde^2$ is cancelled in the action, we can replace it with the
square of the Planck length $l_p^2$ of $M$-theory. Let us see it more
precisely and show the equivalence of the $\widetilde{IIA}$ action to
the $M$-theory action. Concentrating on the Einstein-Hilbert term in the action, the supergravity action for $\widetilde{IIA}$-theory is given by 
\begin{equation}
S^{\widetilde{IIA}} = {1 \over \lstilde^8}\int d^{10}x
           \sqrt{g^{(s)}}e^{-2\phi}R^{(s)}
     = {1 \over 2\pi\lstilde^9}\int_{0}^{2\pi\lstilde}dx^{11}\int d^{10}x
           \sqrt{g^{(s)}}e^{-2\phi}R^{(s)},
\label{eqn:IIAaction}
\end{equation}
where the superscript $(s)$ denotes the string metric of (\ref{eqn:D0D4}).

\noindent
The action (\ref{eqn:IIAaction}) can be rewritten as
\begin{equation}
S^{\widetilde{IIA}} = {1 \over 2\pi\lstilde^9}\int_{0}^{2\pi}dx^{-}
            \int d^{10}x \lstilde\sqrt{g^{(s)}}e^{-2\phi}R^{(s)}
        = {1 \over 2\pi\lstilde^9}\int_{0}^{2\pi}dx^{-}
           \int d^{10}x \sqrt{g^{(11)}}R^{(11)},
\label{eqn:action2}
\end{equation}
where the superscript $(11)$ indicates the uplifted eleven dimensional metric of (\ref{eqn:upM5}).

\noindent
Now it is clear from the above eq. (\ref{eqn:action2}) that the
overall coefficient $\lstilde^2$ of the metric (\ref{eqn:upM5}) is
cancelled by the constant $1/\lstilde^9$ in front of the action. Thus
the r.h.s of the eq. (\ref{eqn:action2}) is equal to the $M$-theory action $S^{M}$, that is, $S^{\widetilde{IIA}} = S^{M}$. 
\parmedskip
Some remarks are in order: 

\noindent
(i) Scrutinizing the uplifted metric (\ref{eqn:upM5}), we can find that the time $t$ in $\widetilde{IIA}$-theory gets mapped to the light-cone time $\tau$ in $M$-theory. This is favorable for the DLCQ interpretaion of the matrix model of M-theory. The crucial point here is that we first take the near-horizon limit and then uplift the metric to eleven dimensions, but these two procedures do not commute, i.e., near-horizon limit + uplifting  $\ne$ uplifting + near-horizon limit. If we adopt the opposite procedure, the time $t$ in $\widetilde{IIA}$-theory becomes the usual time, not the light-cone time, in $M$-theory as well.

\parmedskipn
(ii) We dropped the constant portion $1/\gstilde$ of the gauge
potential $A_0$ in eq. (\ref{eqn:D0D4}). This reflects in the
identification of $x^{11}/\lstilde$ in $\widetilde{IIA}$-theory with
$x^{-}$ in $M$-theory. However it is natural to add it before taking
the near-horizon limit , since the gauge potential $A_0$ goes to zero
at infinity in its presence. On the other hand, after taking the
near-horizon limit, we should drop it away because $1/\gstilde$ is
divergent in this limit. Furthermore it is merely a choice of the gauge whether we add the constant portion or not. Thus we consider it is reasonable to drop $1/\gstilde$ from $A_0$ in the present case.

\parmedskipn
(iii) The light-cone coordinate $x^{-} (= x^{11}/\lstilde)$ in $M$-theory is finite, as it should be. The range of $x^{-}$ is from $0$ to $2\pi$, as we can see it from the explanation for the equivalence of two actions $S^{\widetilde{IIA}}$ and $S^{M}$.

\parmedskipn
(iv) The boundary lies at $\Utilde = \infty$ for the D0 + D4 system,
and $U_{M5} = \infty$ for the \lq\lq boosted" M5-branes,
respectively. Note that the boundary of the D0 + D4 system corresponds
to that of the \lq\lq boosted" M5-branes from the relation $\Utilde =
R^{-1}U_{M5}^2$. This is desirable for a \lq derivation' of matrix
models of M5-branes, or (2,0) superconfomal field theories, as can be
seen from fig. \ref{fig:diagram}. Note also that the effect of the boost, ${Q_0 Q_4^{-1/3} \over R^4U_{M5}^4}(dx^{-})^2$, in the uplifted metric (\ref{eqn:upM5}) becomes negligible approaching to the boundary. Thus this metric takes the form of $AdS_7\times S^4$ in the light-cone coordinate at the boundary. This point will be stressed in the computation of two-point functions in the next section.
%



\subsection{Conformal symmetry}\label{subsec:confsym}

In the previous section, we showed that there exists a conformal symmetry (\ref{e:D0D4translation}) in the D0 + D4 system. Here we re-examine it and will give an alternative represention for that symmetry, which is valid only at the origin of the transverse space. This representation seems to be suitable for identifying the conformal group of the D0 + D4 system with a certain subgroup in a six dimensional conformal group of M5-branes. 

Now let us rewrite the near-horizon metric (\ref{eqn:D0D4}) of the D0 + D4 bound states as
\begin{eqnarray}
ds_{10}^2 &=& \lstilde^2\left[-\left({R \over \sqrt{Q_0}}\right){\Utilde^3 \over \sqrt{Q_4}}dt^2 + \left({\sqrt{Q_0} \over R}\right){\sqrt{Q_4} \over \Utilde^3}(d\Utilde^2 + \Utilde^2 d\Omega_4^2)\right. \nn\\
          && + \left. \sqrt{{Q_0 \over Q_4}}\left\{(d\xdoubtilde^1)^2 
            + (d\xdoubtilde^2)^2 + (d\xdoubtilde^3)^2 + (d\xdoubtilde^4)^2 
            \right\}\right], \nn\\
\label{eqn:rewrittenD0D4}
e^{-2\phi} &=& \left({R \over \sqrt{Q_0}}\right)^3 
               Q_4^{1 \over 2}\Utilde^3,
\qquad A_0 = -\left({R \over \sqrt{Q_0}}\right)^2 \Utilde^3 
              \lstilde.
\end{eqnarray}
Note that the dependence on $R$ ,or equivalently on $\gstilde$, appears only through the combination $R/\sqrt{Q_0}$, except for the transverse part of the metric. Therefore if we restrict ourselves at the origin of the transverse space, $\xdoubtilde^i = 0\; (i=1,2,3,4)$\footnote{Strictly speaking, it is allowed that we are on a three-sphere, $(\xdoubtilde^1)^2 + (\xdoubtilde^2)^2 + (\xdoubtilde^3)^2 + (\xdoubtilde^4)^2 = c^2$, where $c$ is a constant. We can, however, shrink the radius $c$ of the $3$-sphere to zero by a dilatation.}, the transformation laws for $\gstilde$ in (\ref{e:D0D4translation}) can be replaced with those for $\sqrt{Q_0}$. Focusing on the special conformal transformations, we obtain as an alternative representation,
\begin{eqnarray}
\delta_K t &=& -( t^2 + {Q_0Q_4 \over 2R^2\Utilde^4} ), \nn\\
\delta_K \Utilde &=& 2t\Utilde, \nn\\
\label{eqn:newconf}
\delta_K \sqrt{Q_0} &=& 2t\sqrt{Q_0}, \\
\delta_K R &=& 0,\qquad(\mbox{or equivalently},\quad \delta_K \gstilde = 0), 
                \nn\\
\delta_K \xdoubtilde^i &=& -t\xdoubtilde^i \quad(= 0). \nn
\end{eqnarray}
We would like to remark that although the above symmetry is valid only at the origin of the transverse space, this feature matches with the observation for the DLCQ of (2,0) superconfomal field theories \cite{ABS}, in which it was argued that the states invariant under the special conformal symmetry of the quantum mechanics must be concentrated completely at the origin of the moduli space.

Next we read off the transformation laws for the coordinates in $M$-theory from the above transformation. The transformation law for $x^{-} = x^{11}/\lstilde$ is not determined a priori. So we require the conformal invariance, not only for the ten dimensional metric (\ref{eqn:rewrittenD0D4}), but also for the uplifted metric (\ref{eqn:upD0D4}). Then we find\footnote{It is possible to add an arbitrary constant to this transformation law. As can be seen from eq. (\ref{eqn:confsub}) below, the value of the constant should be set to $c^2/2R$, where $c$ is the radius of $3$-sphere mentioned in the last footnote.}
\begin{equation}
\delta_K x^{-} = \delta_K \left({x^{11} \over \lstilde}\right)
               = {2Q_4 \over \Utilde}.
\label{eqn:x11trf}
\end{equation}
Note that the term $Q_0Q_4/2R^2\Utilde^4$ in $\delta_K t$ is crucial to obtain this transformation law.

\noindent 
Using the relations between the coordinates in $\widetilde{IIA}$- and $M$-theories, we get
\begin{eqnarray}
\delta_K X^{+} &=& -\left( (X^{+})^2 + {Q_0Q_4 \over 2R^6U_{M5}^8} \right), 
           \nn\\
\delta_K X^{-} &=& {2Q_4 \over U_{M5}^2}, \nn\\
\label{eqn:Mconf}
\delta_K U_{M5} &=& X^{+}U_{M5}, \\
\delta_K x^i &=& -X^{+}x^i \quad(= 0), \nn
\end{eqnarray}
where we redefined the light-cone coordinates $\tau$ and $x^{-}$ by $X^{+} = \tau/R$ and $X^{-} = Rx^{-}$, respectively. Here we omitted the transformation law for $Q_0$, although it is necessary for the invariance of the eleven dimensional metric under this special conformal transformation. The number of D0-branes, $Q_0$, however, appears only in the term ${Q_0Q_4^{-1/3} \over 2R^6U_{M5}^4}(dX^{-})^2$ in the eleven dimensional metric, and thus it is negligible for large $U_{M5}$.

Now we can see that this conformal symmetry is indeed a subgroup of the six dimensional conformal group of M5-branes, for large $U_{M5}$. The special conformal transformation of M5-branes is given in \cite{KKR},
\begin{eqnarray}
\delta X^{\alpha} &=& \epsilon^{\beta}X_{\beta}X^{\alpha}
             -\epsilon^{\alpha}\left(X^2 + {4\pi N_5 \over U^2}\right)/2,
            \nn\\
\label{eqn:M5SPC}
\delta U &=& -\epsilon^{\alpha}X_{\alpha}U,
\end{eqnarray}
where $\alpha$ and $\beta$ run from $0$ to $5$, and $N_5$ is the
number of M5-branes.

\noindent
Let us look at the special conformal transformation with respect to the light-cone time direction, that is, only $\epsilon^{-}$ is nonvanishing in the above transformation:
\begin{eqnarray}
\delta X^{+} &=& \epsilon^{-}(X^{+})^2, \nn\\
\delta X^{-} &=& -\epsilon^{-}\left({1 \over 2}(X^i)^2 + {2\pi N_5 \over U^2}
                 \right), \nn\\
\label{eqn:confsub}
\delta U &=& -\epsilon^{-}X^{+}U, \\
\delta X^i &=& \epsilon^{-}X^{+}X^i, \nn
\end{eqnarray}
where we define $X^{\pm} = {1 \over \sqrt{2}}\left(X^5 \pm X^0 \right)$, and $i=1,2,3,4$.

When $X^i = 0$, the special conformal symmetry (\ref{eqn:confsub}) of M5-branes coincides with that obtained from the D0 + D4 system (\ref{eqn:Mconf}) under the identification $U = U_{M5}$, $X^i = x^i$, and $\pi N_5 = Q_4$, up to the term $Q_0Q_4/2R^6U_{M5}^8$ in (\ref{eqn:Mconf}) and an irrelevant overall minus sign. For large $U_{M5}$, $Q_0Q_4/2R^6U_{M5}^8$ is much smaller than the other terms, in particular, $2Q_4/U_{M5}^2$. Hence we conclude that the conformal group of the D0 + D4 system can be interpreted as a subgroup of the six dimensional conformal group of M5-branes at the origin of the transverse space, for large $U_{M5}$, i.e., approaching to the boundary. Again this result accords with an observation for matrix models of (2,0) superconformal field theories \cite{ABS}. 



\section{Two-point functions}\label{sec:2ptfunc}

We discussed in detail the compatibility of the scaling limit of Seiberg and Sen with that of Maldacena in the last section. In particular we gave an argument which showed the equivalence of the $\widetilde{IIA}$ action (corresponding to the matrix model) with the $M$-theory action (corresponding to the DLCQ) in the case of (2,0) superconformal field theories. In this section we compute the correlation functions of the DLCQ, or equivalently matrix models, of (2,0) superconfomal theories, by using the correspondence between supergravities and boundary field theories \cite{GKP,wit2pt}. We adopt, in particular, the strategy of \cite{GKP} to the case of the near-horizon geometry of \lq\lq boosted" M5-branes. As we will show below, the results agree with those obtained from matrix models of (2,0) superconformal theories \cite{ABS}. This agreement can be thought of as an evidence either for the Maldacena's conjecture of the correspondence between supergravities and boundary field theories, or for the M(atrix) conjecture especially in the large $N$ limit.

\parmedskip
Let us begin with a brief review of the AdS/CFT correspondence \cite{malda1}\cite{GKP,wit2pt}. It is instructive to consider the $\mbox{AdS}_5/\mbox{CFT}_4$ correspondence as an example. 

At low energy limit, string theory on D3-brane backgrounds reduces to type IIB supergravity in the background of black 3-brane solutions. On the other hand, the low energy effective theory of $N$ D3-branes is described by ${\cal N}=4$ $U(N)$ super Yang-Mills theory on the four dimensional world volume \cite{TASI}. The near-horizon limit of Maldacena \cite{malda1} is just the low energy limit of D3-branes, in which D3-branes decouple from the bulk and the energy of open strings stretched between D3-branes is kept fixed. Thus we can expect that ${\cal N}=4$ $D=4$ $U(N)$ SYM theory is described by type IIB supergravity on the near-horizon geometry of D3-branes, i.e., $AdS_5\times S^5$. In other words,  type IIB SUGRA on $AdS_5\times S^5$ is a sort of the master field theory of ${\cal N}=4$ $D=4$ $U(N)$ SYM theory. In particular, since the loop expansion parameter of type IIB SUGRA on $AdS_5\times S^5$ is of order $1/N^2$, the \lq\lq master field theory" of ${\cal N}=4$ $U(N)$ SYM theory is expected to be the classical type IIB SUGRA on $AdS_5\times S^5$, in the large $N$ limit.

Now let us see the more precise correspondence between type IIB SUGRA on $AdS_5\times S^5$ and ${\cal N}=4$ SYM theory. The former has an $SO(2,4)\times SU(4)$ symmetry, since the $\mbox{AdS}_5$ space has an $SO(2,4)$ isometry which amounts to the four dimensional conformal group at the boundary, and a $5$-sphere has an isometry $SO(6) \sim SU(4)$. On the other hand, the latter becomes a conformal field theory at the origin of the moduli space, and has an $SU(4)$ $R$-symmetry. Therefore the conformal point of ${\cal N}=4$ SYM is conjectured to live at the boundary of the $\mbox{AdS}_5$ space.

Let us proceed to the correspondence of correlation funtions. Since a local operator in field theory is a small disturbance in spacetime, it seems likely that a small fluctuation around the AdS space corresponds to a local operator in conformal field theory at the boundary of the AdS space. To be more precise, the fluctuation around the AdS space must be on-shell in order to pick out only the boundary contribution. From a perspective of string theory, the restriction on the on-shell fluctuations is nothing other than the condition that the beta-function is vanishing, i.e., conformal invariance of the worldsheet. Thus it is natural to consider only the on-shell fluctuations around the AdS space. 

Now let $\phi$ be an on-shell fluctuation. Type IIB SUGRA action $I_{SG}[\phi]$ is likely to accord with the generating functional for connected correlation functions of the boundary conformal field theory. This is because it is the generating functional for connected correlation functions of type IIB closed strings (in $t$-channel), and using the $s$-$t$ channel duality, we can consider it as that of open stings (in $s$-channel). Thus we can arrive at the conjecture of \cite{GKP,wit2pt},
\begin{equation}
\left\langle\exp\int_{{\cal M}_4}\phi_0{\cal O}
    \right\rangle_{\mbox{\scriptsize{CFT}}}
     = \exp\left(iK_{SG}[\phi_0]\right),
\label{eqn:ansatz}
\end{equation}
where ${\cal M}_4$ is the conformal compactification of Minkowski space, ${\cal O}$ is a conformal field, and $K_{SG}[\phi_0]$ is the minimum of $I_{SG}[\phi]$ with the boundary condition, $\phi |_{\mbox{\scriptsize{boundary}}} = \phi_0$.

\noindent
This conjecture is extended to more general cases, i.e., the $\mbox{AdS}_{d+1}/\mbox{CFT}_d$ correspondence, although many of them are not based on string theory.
\parbigskip
Next we turn to the computation of two-point functions of the DLCQ of
(2,0) superconformal theories. We use the uplifted eleven dimensional
metric (\ref{eqn:upM5}), and thus we do not work on the AdS space
exactly. As mentioned in the last section, the uplifted metric
(\ref{eqn:upM5}), however, takes the form of $AdS_7\times S^4$ in the
light-cone coordinate, as we approach to the boundary $U_{M5} =
\infty$. Hence we can expect that our computation will essentially
reduce to that of $AdS_7\times S^4$, and indeed it is as we will see below.



\subsection{Scalars}\label{subsec:scalars}

Now we calculate two-point functions of scalars. The scalar modes around the \lq\lq boosted" M5-branes can be analyzed just in the same way as done completely  by Nieuwenhuizen \cite{Nieuwen}. The only difference appears in the form of Laplacian in the linearized equations.\footnote{This holds for the other modes, and in general the difference arises in the form, not only of Laplacian, but also of differential operators of other kinds.} The reason is the following. The \lq\lq boosted" M5-brane has the spacetime structure ${\cal M}_7\times S^4$, in which ${\cal M}_7$ is an Einstein space with the Ricci tensor $R^{(7)}_{\alpha\beta} = -{3 \over 2Q_4^{2/3}}g^{(7)}_{\alpha\beta}$, and the four form field strength $F_4$ takes the form of the Freund-Rubin ansatz, $F_4 = -3Q_4\epsilon_4$, where $\epsilon_4$ is the volume form of $4$-sphere. These properties are exactly same as those of $AdS_7\times S^4$. Hence the Kaluza-Klein mass spectrum of fluctuation modes coincides with that listed in Table 1 of Ref. \cite{Nieuwen}.

Let us denote scalar modes by $\varphi$ collectively. The linearized equation of scalars is given by
\begin{equation}
\left( \square_7 - M^2\right)\varphi = 0,
\label{eqn:KGeq}
\end{equation}
where the Kaluza-Klein mass $M^2$ can take three series of values, $(k+3)(k+6)$, $(k+2)(k+5)$, $(k=0,1,\cdots)$, and $k(k-3)$, $(k=1,2,\cdots)$, multiplied by a factor $Q_4^{-2/3}$, inverse square of the radius of $S^4$. The seven dimensional Laplacian $\square_7$ is explicitly written as
\begin{eqnarray}
\square_7 = Q_4^{-2/3}\left[ {1 \over 4}{1 \over U_{M5}^5}\del_{U_{M5}}
            U_{M5}^7\del_{U_{M5}} - {Q_0 Q_4 \over R^4 U_{M5}^8}\del_{\tau}^2
            + {2Q_4 \over U_{M5}^2}\del_{\tau}\del_{-}
            + {Q_4 \over U_{M5}^2}\sum_{i=1}^4 \del_i^2\right].
\label{eqn:Laplace}
\end{eqnarray}
For convenience, we introduce a coordinate $z = 1/U_{M5}$. Then the linearized equation (\ref{eqn:KGeq}) is rewritten as
\begin{eqnarray}
\left[ z^5\del_z z^{-5}\del_z - {4Q_0 Q_4 \over R^4}z^6\del_{\tau}^2
            + 8Q_4\del_{\tau}\del_{-}
            + 4Q_4\sum_{i=1}^4 \del_i^2 - {4m^2 \over z^2}\right]
            \varphi = 0.
\label{eqn:rewriteKGeq}
\end{eqnarray}
Here we defined $m^2 = M^2 Q_4^{2/3}$, and it takes three series of integral values, as mentioned above.

Now we expand the scalar $\varphi$ into Fourier modes,
\begin{equation}
\varphi = \varphi (z)\exp (iN_0 x^{-})\int \frac{d\omega}{\sqrt{2\pi}}
          \int \frac{d^4 k}{(2\pi)^2}
           {\hat \varphi}_0(\omega,k^i)\exp 
            (i\omega\tau + i\sum_{j=1}^4 k_j x^j),
\label{eqn:Fourier}
\end{equation}
where $N_0$ is an integer, for the range of $x^{-}$ is from $0$ to $2\pi$. Since we are considering the discrete light-cone quantization of (2,0) superconformal theories, we should not sum over integer $N_0$.

\noindent
As a result, we have for the scalar modes,
\begin{equation}
 \left[ w^5{d \over dw} w^{-5}{d \over dw} 
           + {Q_0\omega^2 \over 64Q_4^3 R^4 (2N_0\omega + k^2)^4}w^6
            \pm 1 - {4m^2 \over w^2}\right]
            \varphi (w) = 0,
\label{eqn:scaleq}
\end{equation}
with $k^2 = \sum_{i=1}^4 k_i^2$, and we rescaled the argument $z$ to $w = 2\sqrt{Q_4 |2N_0\omega + k^2|}z$. Two cases, $\pm$, in the above equation correspond to $2N_0\omega + k^2 < 0$ and $2N_0\omega + k^2 \geq 0$, respectively. 

\parmedskip
A remark is in order:

\noindent
The light-cone momentum $N_0$ naively seems to be equal to the number
of D0-branes, $Q_0$, in view of the DLCQ interpretation of the D0 + D4 system. We should, however, take into account the normalization of the target space coordinates of the quantum mechanics on the instanton moduli space \cite{ABS}. Focusing on the center of mass coordinates, the sigma model action is given by
\begin{equation}
S = {1 \over \gstilde\lstilde}\int dt
                 \del_t \xtilde^i\del_t \xtilde^i
 = {\lstilde \over \gstilde}\int dt
                 \del_t \xdoubtilde^i\del_t \xdoubtilde^i
 = R\int dt
    \del_t \xdoubtilde^i\del_t \xdoubtilde^i
 = {1 \over R}\int dt
    \del_t x^i\del_t x^i,
\label{eqn:SigmaModel}
\end{equation}
where all the quantities were defined in the last section. Those with
and without tilde are associated with the near-horizon geometry of the
D0 + D4 system and the \lq\lq boosted" M5-branes, respectively. The index $i$ runs from $1$ to $4$.

\noindent
Thus the center of mass part of the Hamiltonian is proportional to $R$, not to $R/Q_0$, although the latter is the natural normalization for the DLCQ interpretation. This indicates that the dependence on $Q_0$ is absorbed into the coordinates $x^i$, and we should set $N_0$ to $1$, in our normalization.

Now we solve the linearized eq. (\ref{eqn:scaleq}) for scalars. Although it is not easy to solve it exactly, we only need the asymptotic behavior of the solutions in the vicinity of the boundary of the spacetime, i.e., $w=0$ ($U_{M5} = \infty$) and $w=\infty$ ($U_{M5} = 0$), for our purpose. 

Suppose that $Q_0 \sim Q_4$, denoting them by $Q$, and $\omega \sim k^2$.\footnote{$\omega$ has mass dimension 2, since the light-cone time $\tau = Rt$ has dimension $(\mbox{length})^2$.} Then the differential eq. (\ref{eqn:scaleq}) is schematically rewritten as 
\begin{equation}
\left[ w^5{d \over dw} w^{-5}{d \over dw} 
           + {w^6 \over (QR^2 \omega)^2}
            \pm 1 - {4m^2 \over w^2}\right]
            \varphi (w) = 0.
\label{eqn:schem}
\end{equation}
We further assume that $|QR^2 \omega| \ll 1$.\footnote{We would like to emphasize that this condition for $\omega$ (and so $k^2$) is not too restrictive, since we are interested only in the behavior of correlation functions at long distance.} Let us divide this equation into three regions. They are characterized by (I) $|w^4/QR^2 \omega| \ll 1$, (II) $|w| \ll 1$ and $|w^3/QR^2 \omega| \gg 1$, and (III) $|w^4/QR^2 \omega| \gg 1$, respectively. Accordingly the differential eq. (\ref{eqn:schem}) takes the form in each region,
\begin{eqnarray}
\mbox{(I)}&&\quad \left[ w^5{d \over dw} w^{-5}{d \over dw} 
            - {4m^2 \over w^2}\right]\varphi (w) = 0, 
\label{eqn:difeqdiv1}\\
\mbox{(II)}&&\quad \left[ w^5{d \over dw} w^{-5}{d \over dw} 
           + {w^6 \over (QR^2 \omega)^2}
           - {4m^2 \over w^2}\right]\varphi (w) = 0, 
\label{eqn:difeqdiv2}\\
\mbox{(III)}&&\quad \left[ w^5{d \over dw} w^{-5}{d \over dw} 
           + {w^6 \over (QR^2 \omega)^2}
           \right]\varphi (w) = 0,
\label{eqn:difeqdiv3}
\end{eqnarray}

\noindent
As we will show in more detail in the Appendix B, we can find two independent solutions for the differential equation (\ref{eqn:scaleq}). Matching the solutions in the overlap regions, they behave like, in each region,
\begin{equation}
\varphi (w) = 
\left\{ 
\begin{array}{ll}
\!\!\!\mbox{(I)}&\!\!\!
  {2^{\nu/4}(4QR^2\omega)^{(\nu -3)/4} \over \Gamma(-\nu/4 +1)}w^{3-\nu},\\
\!\!\!\mbox{(II)}&\!\!\!
 y^{3/4} J_{-\nu /4}(y),\\
\!\!\!\mbox{(III)}&\!\!\!
\sqrt{{2 \over \pi}}y^{1/4}
              \cos\left(y+{(\nu/2 - 1)\pi \over 4}\right),
\end{array}
\right.
\mbox{and}
\left\{
\begin{array}{rl}
&\!\!\!\!\!\!
  {2^{-\nu/4}(4QR^2\omega)^{-(\nu +3)/4} \over \Gamma(\nu/4 +1)}
      w^{3+\nu},\\
&\!\!\!\!\!\!y^{3/4} J_{\nu /4}(y),\\
&\!\!\!\!\!\!\sqrt{{2 \over \pi}}y^{1/4}
              \cos\left(y-{(\nu/2 + 1)\pi \over 4}\right),
\end{array}
\right.
\label{eqn:indepsol}
\end{equation}
where $\nu = \sqrt{4m^2 + 9}$ and takes odd integral values $2k+9,2k+7$ ($k=0,1,\cdots$) and $2k-3$ ($k=1,2,\cdots$). $J_{\pm\nu/4}$ are the Bessel functions, and $y=w^4/4QR^2 \omega$ (strictly $\sqrt{Q_0Q_4}|\omega|w^4/2^5R^2Q_4^2|2N_0\omega + k^2|^2$).

In order to compute two-point functions, it is necessary to take into account sub-leading contributions to the first solution in the region (I). The region near the boundary, $w=0$ ($U_{M5}=\infty$), is characterized by $|w^2/QR^2\omega| \ll 1$ ($z/R \ll 1$), and lies within the region (I). Since $|w^6/QR^2\omega| \ll 1 \ll |1/w^2|$ in this region, we can consider the following differential equation to find sub-leading contributions:
\begin{eqnarray}
\left[ w^5{d \over dw} w^{-5}{d \over dw} 
           - {4m^2 \over w^2}\right]
            \varphi (w) = \mp\varphi (w),
\label{eqn:w0region}
\end{eqnarray}
where we treat the r.h.s as a perturbative correction to eq. (\ref{eqn:difeqdiv1}) in the region (I). Note that this equation is nothing but the Bessel (or modified Bessel) equation for $2N_0\omega + k^2 < 0$ (or $2N_0\omega + k^2 \geq 0$), and is exactly same as the one for massive scalars in $AdS_7\times S^4$ \cite{wit2pt}.\footnote{Strictly speaking, we are working in the Lorentzian, not in the Euclidean, space.\cite{BKLT}} 

\noindent
Thus we obtain sub-leading contributions to the first solution, which we will denote $\varphi_1 (w)$, near the boundary $w=0$ as
\begin{eqnarray}
\varphi_1 (w) &=& {2^{\nu/4}(4QR^2\omega)^{(\nu -3)/4} \over \Gamma(-\nu/4 +1)}
  \left[w^{3-\nu} + \sum_{n=1}^{\nu -1}(i)^{n \mp n}
  {(\nu -n-1)! \over 2^{2n}n!(\nu -1)!}w^{3+2n-\nu}\right. \nn\\
  &&\left. \qquad\qquad
     \mp {(i)^{(\nu -1) \mp (\nu -1)} \over 2^{2\nu -1}\nu [(\nu -1)!]^2}
   \left(w^{3+\nu}\log w - {1 \over 2\nu}w^{3+\nu}\right)
  +\cdots\right].
\label{eqn:sol1corr}
\end{eqnarray}
Here the dots \lq$\cdots$' denote the higher order corrections, and they are irrelevant to the computation of two-point functions. Note that $\nu$ is an odd integer, as mentioned previously.

Now it is crucial to determine the boundary conditions for scalars in the calculation of two-point functions. At the boundary $z/R =0$ ($w=0$), it is determined so that we normalize the boundary value of $\varphi (w)$, regularized by an IR (UV) cutoff $\epsilon$ ($z\geq\epsilon$) on the bulk (boundary) theory, to $1$ \cite{GKP}. We do not, however, have a natural way to fix the boundary condition at $z/R =\infty$ ($y=\infty$), as opposed to the case of the Euclidean AdS space \cite{BKLT}.\footnote{This ambiguity of the boundary condition is similar to that discussed in \cite{BKLT} in the case of the Lorentzian AdS space, but they are in fact different. In our case, the ambiguity comes from the boosted effect, rather than Lorentzian nature of our geometry.} Here we shall choose the boundary condition at $z/R =\infty$ ($y=\infty$) in the following way:

\noindent
Since we are now considering a matrix quantum mechanics, it is usual to perform a Wick rotation $\tau \to -i\tau$, or equivalently $\omega \to i\omega$. Accordingly we should rotate $x^-$ to $ix^-$, or equaivalently $N_0 \to -iN_0$, to keep the metric of the \lq\lq boosted'' M5-branes real valued. We should, however, note that it is somewhat strange to adopt this Wick rotation in view of the DLCQ interpretation, since the time $\tau$ in a matrix quantum mechanics is the light-cone time in DLCQ. But our attitude here is that we can justify this Wick rotation by referring to matrix descriptions of theories in DLCQ, and we do perform the usual Wick rotation for a matrix quantum mechanics. Then we can fix the boundary condition at $z/R =\infty$ ($y=\infty$) by choosing the solution, in the region (III), which falls off exponentially for $z/R \gg 1$ ($y \gg 1$).

\noindent
As a result, we have a solution
\begin{eqnarray}
\varphi (w) = \frac{e^{i(\nu -2)\pi/8}\varphi_1 (w)
                + ie^{-i(\nu -2)\pi/8}\varphi_2 (w)}
              {e^{i(\nu -2)\pi/8}\varphi_1 
           ({\scriptstyle 2\sqrt{Q_4 |2N_0\omega + k^2|}}\epsilon)
                + ie^{-i(\nu -2)\pi/8}\varphi_2 
           ({\scriptstyle 2\sqrt{Q_4 |2N_0\omega + k^2|}}\epsilon)},
\label{eqn:solwbc}
\end{eqnarray}
where $\varphi_1 (w)$ and $\varphi_2 (w)$, respectively, denotes the first and the second solution (including the sub-leading contributions) in eq. (\ref{eqn:indepsol}).

According to the ansatz (\ref{eqn:ansatz}), two-point functions in momentum space are given by
\begin{equation}
\left\langle{\cal O}(\omega,k^2){\cal O}(\omega',k'^2)
    \right\rangle
     = {\delta^2 \over \delta\hat{\varphi}_0 (\omega,k^2)
      \delta\hat{\varphi}_0 (\omega',k'^2)}K_{SG}[\varphi_0],
\label{eqn:2ptmom}
\end{equation}
where $\varphi_0$ is the Fourier transform of $\hat{\varphi}_0$. The minimum $K_{SG}[\varphi_0]$ of the SUGRA action is written as
\begin{equation}
K_{SG}[\varphi_0] = \left. c\frac{\pi^2}{2}Q_4^{4/3}
                \int d^6 x\sqrt{-\det g^{(7)}}
               g^{U_{M5}U_{M5}}
              \varphi\del_{U_{M5}}\varphi \right |_{U_{M5} = 0}^
              {U_{M5} = 1/\epsilon}.
\label{eqn:mimM5act}
\end{equation}
Here $c$ is a normalization constant of the 11D SUGRA action. The factor $\frac{\pi^2}{2}Q_4^{4/3}$ is the volume of 4-sphere.

\noindent
Thus the two-point function for scalars is given by 
\begin{eqnarray}
\left\langle{\cal O}(\omega,k^2){\cal O}(\omega',k'^2)
    \right\rangle
&=& -c{\pi^3 \over 2}(\sqrt{Q_4}\epsilon)^{2\nu -6}\delta(\omega +\omega')
    \delta^4(\vec{k}+\vec{k'}) \nn\\
&\times& Q_4^3
\Biggl[\frac{4}{[(\nu -1)!]^2}
(2N_0\omega +k^2)^{\nu}\log|2N_0\omega +k^2|
\nn\\
&&\qquad - 2\nu
\frac{\Gamma (1-\nu/4)}{\Gamma (1+\nu/4)}
\left(\frac{\sqrt{Q_0/Q_4}|\omega|}{4R^2}\right)^{\nu/2}\Biggr].
\label{eqn:2ptscmom}
\end{eqnarray}
Performing a Fourier transformation to position space, the two-point function becomes
\begin{eqnarray}
\left\langle{\cal O}(x^{+},\vec{x}){\cal O}(0,\vec{x'})
    \right\rangle
&=& -c(\sqrt{Q_4}\epsilon)^{2\nu -6}Q_4^3
     \Biggl[N_0^{\nu}\frac{2^{\nu -2}
     \pi\nu !}{[(\nu -1)!]^2}\frac{1}{(Rx^{+})^{\nu +3}}
     \exp\left(-{iN_0 \over 2}\frac{(\vec{x}-\vec{x'})^2}{Rx^{+}}\right)
\nn\\
&&\!\!\!\!\!\!\!\!
+ 2^{-\nu/2}\pi^{5/2}\nu\frac{\Gamma (1/2 +\nu/4)}{\Gamma (1 +\nu/4)}
  \!\!\left(\frac{\sqrt{Q_0/Q_4}}{R^2}\right)^{\nu/2}\!\!\!\!\!\!
  \frac{1}{(Rx^{+})^{\nu/2 +1}}\delta^4 (\vec{x}-\vec{x'})
  \Biggr],
\label{eqn:2ptscpos}
\end{eqnarray}
where the light-cone time $x^{+}$ is equal to $\tau/R$, and $N_0$ should be $1$, as discussed above. The cutoff $\sqrt{Q_4}\epsilon$, rather than $\epsilon$, is employed in the computation of correlation functions in the AdS/CFT correspondence \cite{FMMR}.

\parmedskip
Some remarks are in order:

\parmedskipn
(i) The two-point function consists of two elements, up to contact terms. The first element in eq. (\ref{eqn:2ptscpos}) is of the expected form. It is exactly the two-point function of scalar primary operators with dimension $\Delta =\nu +3$ in DLCQ of a conformal field theory \cite{ABS}. The dimension $\Delta$ takes values in even integrals, $2k +12$, $2k +10$ ($k=0,1,\cdots$) and $2k$ ($k=1,2,\cdots$), and it coincides with the analysis of \cite{LR,AOY,Min} for (2,0) superconfomal field theories.

\parmedskipn
(ii) The second element in eq. (\ref{eqn:2ptscpos}) has the correct dimension with respect to the conformal symmetry in section \ref{subsec:confsym}. This term is, however, not desirable from the viewpoint of the DLCQ of (2,0) superconformal theories. But as far as the IR behavior of the theory is concerned, we can drop it away because it contains a delta-function factor $\delta^4 (\vec{x}-\vec{x'})$. Thus we have 
\begin{equation}
\left\langle{\cal O}(x^{+},\vec{x}){\cal O}(0,\vec{x'})
    \right\rangle
\sim 
\frac{1}{(Rx^{+})^{\Delta}}
     \exp\left(-{iN_0 \over 2}\frac{(\vec{x}-\vec{x'})^2}{Rx^{+}}\right).
\label{eqn:2ptscIR}
\end{equation}

\parmedskipn
(iii) Having discarded the second term in eq. (\ref{eqn:2ptscpos}), the above equation (\ref{eqn:2ptscIR}) shows the boost invariance of the DLCQ description, since the light-cone time $x^{+}$ and the radius $R$ of the light-like circle appear only through the combination $Rx^{+}$.

\parmedskipn
(iv) In a direct analysis of the quantum mechanics on an instanton moduli space, it is necessary to regularize the singularities corresponding to small instantons. In fact it was done by adding a Fayet-Iliopoulos term to the Yang-Mills quantum mechanics of D0 + D4 bound states \cite{ABS}, and the resolved moduli space of instantons were interpreted as the moduli space of instantons on non-commutative ${\bf R}^4$ \cite{NekSchw}. In the present case we have reguralized the UV divergence of the boundary theory by introducing an IR cutoff $\epsilon$ (or $\sqrt{Q_4}\epsilon$) of the bulk theory. But we are not sure how our regularization scheme is related to that of \cite{ABS,NekSchw}. 



\section{Discussions}\label{sec:discuss}


In the present work we analyzed the large $N$ limit of matrix models of M5-branes, or (2,0) superconformal field theories, by applying the Maldacena's conjecture of the Bulk/Boundary correspondence to the DLCQ of M5-branes. 
We discussed in detail the relation between the near-horizon limit and the Seiberg-Sen limit. This analysis seems to support the Seiberg and Sen's argument \cite{seiberg,sen} of a justification for the DLCQ interpretation of the matrix model of M-theory. 

In particular we can interpret the generalized conformal symmetry in D0 + D4 bound states (corresponding to matrix model) as a conformal symmetry of the DLCQ of six-dimensional conformal field theories (corresponding to M-theory). To do so, it turned out that we should consider the transformation of the number of D0-branes $Q_0$, rather than that of the string coupling constant $g_s$ (or $\gstilde$), in the generalized conformal transformation of D0 + D4 bound states.

We also calculate two-point functions of scalars in the DLCQ of (2,0) superconformal theories, by employing the conjecture \cite{GKP,wit2pt} for the AdS/CFT correspondence, although the near-horizon geometry of \lq\lq boosted" M5-branes is not an AdS space. Due to the fact that the near-horizon geometry of \lq\lq boosted" M5-branes is still an Einstein space, the computation \cite{Nieuwen} of Nieuwenhuizen has been carried over in our case. In fact two-point functions contain a subtle contribution, but it can be dropped away as far as the IR behavior of the theory is concerned. Thus we can obtain an expected result. This also shows  an evidence for the Maldacena's conjecture in the case of \lq\lq boosted" $p$-branes \cite{CLP}.

It is obvious that we should make a direct analysis of matrix models of M5-branes and compare it with the results here. In particular it is interesting to see how the generalized conformal symmetry in sect. \ref{subsec:confsym} is realized in the direct analysis of matrix models. It might be related to large $N$ renormalization group transformations of matrix models \cite{HINS} and \cite{douglas,lowe}, and shed some lights on how to determine the large $N$ limit of the matrix model of M-theory.

\section*{Acknowledgements}

We are grateful to T.~Yoneya for useful discussions. We are also indebted to M.~Fukuma, H.~Kunitomo and K.~Murakami for discussions. S.~H. would like to thank M.~Kato, Y.~Kazama, J.~Polchinski, S.-J.~Rey and H.~S.~Yang for comments. The work of S.~H. was supported in part by the Japan Society for the Promotion of Science.





\section*{Appendix A}\label{sec:appenA}


We extend some results in sect. \ref{sec:branesol}, and in particular present generalized conformal symmetries, probe actions and the Routhians in more general cases.
The supergravity solution corresponding to $N_p$ coincident D$p$-branes is
\bea
ds_{10}^2 &=&
H_p^{-{1\over 2}}(-dt^2 + dx_{\parallel}^2 )
+H_p^{1\over 2} (dr^2 + r^2 d\Omega_{8-p}^2),
\cr
e^{-2\phi} &=& 
g_s^{-2} H_p^{{p-3\over 2}},
\qquad\qquad
H_p = 1+{g_s Q_p l_s^{7-p} \over r^{7-p}},
\cr
A_{01\cdots p} &=& 
g_s^{-1} (1-H_p^{-1}),
\label{e:Dp}
\eea
where $x_{\parallel}$ collectively denotes the coordinates parallel to D$p$-branes, and $r$ and $\Omega_{8-p}$ are the polar coordinates transverse to D$p$-branes. The parameter $Q_p$ is proportional to the number $N_p$ of D$p$-branes.
In the near-horizon limit, $r\to 0$ and 
$l_s\rightarrow 0$, keeping fixed 
$U\equiv {r/ l_s^2}$ and 
$g_{YM}^2 = g_s l_s^{p-3}$,
the above solution becomes
\bea
ds_{10}^2 &=&
l_s^2\left[
H^{-1} U^2 (-dt^2 + dx_{\parallel}^2) 
+ H (U^{-2} dU^2 + d\Omega_{8-p}^2)
\right],
\cr
e^{-2\phi} &=& 
Q_p^2 H^{p-7},
\qquad\qquad
H = { g_{YM}^2 \sqrt{Q_p} \over U^{3-p\over 2} }
\cr
A_{01\cdots p} &=& 
- Q_p (l_s U)^{p+1} H^{-4}.
\label{e:NearDp}
\eea
Here we dropped a constant part of $A_{01\cdots p}$ with a suitable choice of the gauge.
For $p=3$, $H$ is a constant in $U$ and the 
space-time is $AdS_5 \times S^5$ with radius $l_s\sqrt H$.

Now let us look at the transverse part of the metric.
\bea
ds_{10}^2 &\propto &
- H^{-1}U^2 dt^2 + H\left(U^{-2}dU^2 + d\Omega^2\right),
\cr
e^{-2\phi} &\propto & H^\gamma,
\qquad\qquad
H = 
\left({g \over U^{\alpha}}\right)^\beta,
\label{e:GeneralMetric}
\eea
where $g$ is proportional to $g_s$.

\noindent
This is in general invariant under the following $SU(1,1)$ transformation
\be
\begin{array}{rcl}
\delta_H\, t   &=& 1, \\
\delta_H\, U   &=& 0, \\
\delta_H\, g   &=& 0, 
\end{array}
\qquad
\begin{array}{rcl}
\delta_D\, t   &=& -t, \\
\delta_D\, U   &=& U, \\
\delta_D\, g   &=& \alpha g, 
\end{array}
\qquad
\begin{array}{rcl}
\delta_K\, t   &=& -t^2- (\alpha \beta +1)^{-1} H^2 U^{-2}, \\
\delta_K\, U   &=& 2tU, \\
\delta_K\, g   &=& 2\alpha tg.
\end{array}
\label{e:Dptranslation}
\ee
In the case of D$p$-branes at $x_{\parallel}=0$ and of D0 + D4 bound states,
a set of parameters $(\alpha, \beta, \gamma)$ is equal to 
$(3-p,{1/2},p-7)$ and $(1,1,-3)$, respectively. 
The effective action for a probe particle moving along the radial direction $U$ in the background of source $N_p$ D$p$-branes 
is determined by this $SU(1,1)$ symmetry as
\be
S_{eff} =
\int dt\, 
\left(1-\sqrt{1-w}\right)f\left(H\right) U,
\qquad
w = {H^2 \over U^4} \dot U^2
  = \left({g^\beta \dot U\over U^{\alpha \beta +2}}\right)^2.
\label{e:Dpprobe}
\ee
with an arbitrary function $f$.


\noindent
In the case of pure D0-branes and D0 + D4 bound states, the eleven dimensional metric and the gauge potential $A_0$ take the form
\bea
ds_{11}^2 &=&
e^{r\phi} ds_{10}^2 +
e^{s\phi} (dx^{11} - A_0 dt)^2,
\cr 
&=&
-2 e^{s\phi} A_0 dt dx^{11} + 
   e^{s\phi}(dx^{11})^2 + 
   e^{r\phi}g_{ij}^{(10)} dx^i dx^j,
\cr
A_0^2 
&=& 
- e^{(r-s)\phi} g_{tt}^{(10)},
\eea
with $r=-2/3$ and $s=4/3$.
Thus $t$ and $x^{11}$ correspond to the light-cone coordinates $x^\pm$.
Then a particle action with fixed light-cone momentum $p_-$ is described by the Routhian,
\be
dt\, {\cal R} =
dt\,  p_- A_0 
\left(
1-\sqrt{1 + { g_{ij}^{(10)}\over g_{tt}^{(10)} } \dot x^i \dot x^j }
\right).
\ee
If $\dot{\Omega} = 0$ (and $\dot{x}_{\parallel}=0$), this coincides with that of eq.\ \eq{e:Dpprobe}
with $f(H)U = p _- A_0$.




\section*{Appendix B}\label{sec:appenB}

In this appendix, we give in more detail a construction of the solution for scalars in section \ref{subsec:scalars}.

We can easily find the solutions for each region. In region (I), two independent solutions are given by $w^{3\pm\nu}$ with $\nu =\sqrt{4m^2 +9}$.
We can rewrite the equations for region (II) and (III), respectively, into 
\begin{eqnarray}
\mbox{(II)}&&\quad 
\left[ y^{1/2}{d \over dy}y^{-1/2}{d \over dy} + 1 - {m^2 \over 4y^2}
\right]\varphi (y) = 0, \\
\mbox{(III)}&&\quad 
\left[ y^{1/2}{d \over dy}y^{-1/2}{d \over dy} + 1 
\right]\varphi (y) = 0,
\label{eqn:eqforII}
\end{eqnarray}
where $y$ is schematically equal to $w^4/4QR^2 \omega$, and strictly to $\sqrt{Q_0Q_4}|\omega|w^4/2^5R^2Q_4^2|2N_0\omega + k^2|^2$.
Thus, in region (II), we have the solutions $y^{3/4}J_{\pm\nu/4}(y)$, and those in region (III) are given by their asymptotic forms $\sqrt{{2 \over \pi}}y^{1/4}\cos\left(y\mp{(\nu/2 \pm 1)\pi \over 4}\right)$ at $y\to\infty$.

Next let us match the solutions in the overlap regions. We can trivially match the solutions in region (II) and (III). We need more care for matching the solutions in region (I) and (II). The overlap region between (I) and (II) are characterized by $|QR^2\omega|^{1/3}\ll |w|\ll |QR^2\omega|^{1/4}$, or equivalently by $1 \ll |w^6/(QR^2\omega)^2| \ll |1/w^2|$. Hence in this overlap region we can obtain the sub-leading contributions to the solutions $w^{3\pm\nu}$ in region (I) from the equation
\begin{eqnarray}
\left[ w^5{d \over dw} w^{-5}{d \over dw} 
          - {4m^2 \over w^2}\right]\varphi (w)
= - {w^6 \over (QR^2 \omega)^2}\varphi (w).
\label{eqn:matchIaII}
\end{eqnarray}
This can be rewritten, in terms of the variable $y(\ll 1)$, as
\begin{eqnarray}
\left[ y^{1/2}{d \over dy}y^{-1/2}{d \over dy} - {m^2 \over 4y^2}
\right]\varphi (y) = -\varphi (y).
\label{eqn:matchy} 
\end{eqnarray}
Here we can consider the r.h.s. as a perturbative correction. Denoting the sub-leading contributions by $\tilde{\varphi}(y)=\tilde{\varphi}_1(y)+\tilde{\varphi}_2(y)+\cdots$, and the homogeneous solutions $y^{(3\pm\nu)/4}(=w^{3\pm\nu}/(4QR^2\omega)^{(3\pm\nu)/4})$ by $\tilde{\varphi}_0(y)$, the sub-leading contributions $\tilde{\varphi}(y)$ are obtained iteratively by
\begin{eqnarray}
\tilde{\varphi}_i(y) = \int_0^y d\xi {2\xi^{-1/2} \over \nu}
               \left(\xi^{(3+\nu)/4}y^{(3-\nu)/4}
                  -\xi^{(3-\nu)/4}y^{(3+\nu)/4}\right)
              \tilde{\varphi}_{i-1}(\xi)
      \qquad (i\geq 1).
\label{eqn:iterate}
\end{eqnarray}
As a result, we have
\begin{eqnarray}
\varphi (y) &=& \tilde{\varphi}_0(y) + \tilde{\varphi}_1(y)+\cdots 
= 2^{\pm\nu/4}\Gamma (\pm\nu/4 + 1)y^{3/4}J_{\pm\nu/4}(y),
\label{eqn:Bessel} 
\end{eqnarray}
and thus the two solutions $w^{3\pm\nu}$ in region (I) are connected to $\frac{\Gamma (\pm\nu/4 + 1)}{2^{\mp\nu/4}(4QR^2\omega)^{-(3\pm\nu)/4}}y^{3/4}J_{\pm\nu/4}(y)$ in region (II) (not a linear combination of them), as given in eq. (\ref{eqn:indepsol}).




\end{document}